# Personalized workflow to identify optimal T-cell epitopes for peptide-based vaccines against COVID-19

Rui Qiao[1], Ngoc Hieu Tran[2], Baozhen Shan[3,*], Ali Ghodsi[1], Ming Li[2,*]

[1] Department of Statistics and Actuarial Science, University of Waterloo, Waterloo, Ontario, Canada.

[2] David R. Cheriton School of Computer Science, University of Waterloo, Waterloo, Ontario, Canada.

[3] Bioinformatics Solutions Inc., Waterloo, Ontario, Canada.

* Corresponding authors.

Emails: mli@uwaterloo.ca (Ming Li); bshan@bioinfor.com (Baozhen Shan).

**Abstract**.

Traditional vaccines against viruses are designed to target their surface proteins, i.e., antigens, which can trigger the immune system to produce specific antibodies to capture and neutralize the viruses. However, viruses often evolve quickly, and their antigens are prone to mutations to avoid recognition by the antibodies (antigenic drift). This limitation of the antibody-mediated immunity could be addressed by the T-cell mediated immunity, which is able to recognize conserved viral HLA peptides presented on virus-infected cells. Thus, by targeting conserved regions on the genome of a virus, T-cell epitope-based vaccines are less subjected to mutations and may work effectively on different strains of the virus. Here we propose a personalized workflow to identify an optimal set of T-cell epitopes based on the HLA alleles and the immunopeptidome of an individual person. Specifically, our workflow trains a machine learning model on the immunopeptidome and then predicts HLA peptides from conserved regions of a virus that are most likely to trigger responses from the person's T cells. We applied the workflow to identify T-cell epitopes for the SARS-COV-2 virus, which has caused the recent COVID-19 pandemic in more than 100 countries across the globe.

## INTRODUCTION

Since Louis Pasteur experimented with live attenuated cholera vaccine and inactivated anthrax vaccine at the end of the 19th century, these methods of producing vaccines have saved billions of lives and are still in use today. Such vaccines are designed to stimulate the host immune system to elicit specific antibodies that bind to the corresponding antigens on the surface of a virus and neutralize the virus. In addition, memory B cells are formed to provide long-term protection by remembering the antigens and quickly producing the antibodies in future re-exposure to the same virus. However, both antibodies and memory B cells are specific to the antigens, and the later are prone to mutations to avoid recognition by the former (so called "antigenic drift"). Once a mutated virus escapes the antibodies and infects the host cells, it now depends on the T-cell mediated immunity to fight against the virus. Inside the infected cells, viral proteins are processed into short peptides, which are then loaded onto MHC proteins (MHC: major histocompatibility complexes). Subsequently, the MHC-peptide complexes are presented on the surface of infected cells for recognition by specific T cells. Activated $CD8^+$ T cells then recognize and clear the infected cells. Similar to the antigen-antibody association, the T-cell immunity also depends strictly on the MHC-peptide complexes. MHC proteins are encoded by human leukocyte antigen (HLA), a gene complex that is located among the most genetically variable regions on the human genome. HLA alleles essentially change from one individual to another, and each allele can only present a certain set of peptides. HLA peptides that can be presented on the surface of infected cells and recognized by T cells are called T-cell epitopes. Last but not least, for a vaccine to provide long-term and robust protection, it is essential to identify T-cell epitopes that originate from conserved regions of the virus. Given the extremely high specificity of the MHC-peptide-T cells mechanism, a personalized approach is desirable to precisely identify optimal conserved T-cell epitopes for each individual person.

As of March 11, 2020, COVID-19 has been spreading to over 100 countries on all continents. Within about three months since December 2019, the number of infected cases has soared to more than 120,000 and the number of deaths has surpassed 4,000 worldwide. While the epidemic initially started in China with majority of infected cases (~70%), massive clusters of thousands of cases have broken out in several other countries, including Italy, Iran, South Korea, France, Spain, Germany, and the United States. The outbreak has also crippled the global economy, causing the worst chaos and market rout since the financial crisis in 2008. A potent vaccine is the only solution to this pandemic.

Traditional vaccine designs may have limitations when applied to COVID-19. While having less fatality rate, the number of infected cases of COVID-19 in just three months has exceeded 10 times the total number of SARS and MERS, two other viruses from the corona family that have caused recent outbreaks in 2003 and 2012. This very high level of contagious may suggest that COVID-19 is able to mutate quickly for adaptation, probably even more than influenza. Thus, vaccines based on antibodies and memory B cells may become obsolete soon after introduced. For instance, in the case of influenza, we have to inoculate people with an updated vaccine every year, which is often only partially effective [1]. Traditional universal vaccines may also cause autoimmune diseases such as multiple sclerosis, type 1 diabetes, systemic lupus [2].

T-cell epitope-based vaccines have been explored in recent years thank to their capability of targeting conserved epitopes and hence providing long-term protection against different strains of viruses [17]. Using epitopes as vaccine ingredients also has the safety advantage than using attenuated or inactivated viruses. T-cell epitope-based vaccines have been developed for malaria, HIV, Hepatitis C virus, influenza, and tuberculosis (BCG), with various success rates [4]. While BCG induces T-cell responses to a large number of antigens, for most diseases, the vaccines depend only on a few selected epitopes [3]. A recent study on influenza by Auladell *et al.* suggests that CD8+ T cell response has an emerging role in influenza protection [14].

To our knowledge, while neoantigens have been used for cancer immunotherapy vaccines [19,20], all prior works on T-cell virus vaccines only involved peptides from the pathogen proteins but not neoantigens that are actually presented by the HLA class I complex (and their subtypes) determined by mass spectrometry. These neoantigens are not universal for all HLA types and are not necessarily obtained directly from the pathogen proteins, as they could be cis- and trans-spliced peptides or from noncoding regions [9,5]. Thus, we propose a personalized workflow to use mass spectrometry together with our recent deep learning models [7,8,10] to directly identify conserved T-cell epitopes on the surface of virus-infected cells.

## RESULTS

### A workflow for the T-cell vaccine design

Figure 1 describes our workflow for HLA-specific anti-viral CD8+ T-cell peptide-based vaccine design. This workflow combines our cancer immunotherapy platform [6] and parts from [16]. The details of each step in the workflow are explained in the following:

1. $10^8$ sick cells are obtained from patient blood sample.
2. Genome sequencing gives 6 HLA types, as well as TCR sequences.
3. Immunopeptidome is obtained by mass spectrometry and analyzed with PEAKS X [10] database search using no-enzyme option. The immunopeptidome is used to re-train DeepNovo [7,8] for each HLA subtype on the left.
4. Personalized DeepNovo model is used to analyze the mass spectrometry data again to find peptides that were missing from the PEAKS X database search (these are *cis*- or *trans*-spliced peptides).
5. For each of the 6 HLA types, this personalized approach enables comprehensive and accurate identification of neoantigens.
6. Effective neoantigens are selected for each HLA allele.

Previous research in SARS-CoV suggests that spike (S) protein is the main antigenic component responsible for inducing the host immune responses. Thus, it is an important target for vaccine and antiviral development [12]. Promising results for vaccines developed based on the full sequence of S protein have been reported on mouse models [13].

Without mass spectrometry data, just using the computational part of the workflow in Figure 1, we can identify potential target peptides for developing an epitope-based vaccine against the nCov-2019. Table 1 reports optimal T-cell epitopes for each of the top 20 most frequent HLA class I alleles in Chinese population [15]. Following [16], we used NetCTLpan 1.1 server to find potent target peptides from the S protein of nCoV-2019 (GenBank: MN908947.3). The NetCTLpan tool integrates peptide MHC binding prediction, proteasomal C terminal cleavage and TAP transport efficiency. A combined rank score is given for each pair of peptide and HLA allele. This score represents the rank of prediction score to a set of 1,000,000 random natural 9-mer peptides. We filtered NetCTLpan-predicted peptides with a rank score threshold of 2%. After filtering, 177 peptides remained, and on average, each allele had 23 candidate peptides. Alignment of those peptides against the Uniprot human database showed that none of them was self-peptide. Next, we applied the IEDB immunogenicity prediction tool to predict the immunogenicity of those 177 peptides. The median of immunogenicity scores was used as a threshold to select the top 50% peptides. Table 1 shows the top 3 candidates and their corresponding predicted immunogenicity score for each included allele. The candidate peptides were ranked by the MHC score predicted by NetCTLpan. As we can see from the results, there are 13 unique peptides among 15 candidate peptides reported for the top 5 most frequent alleles. This shows that different HLA alleles "prefer" different viral peptides with different binding affinities. This fact has three implications:

1. An epitope-based vaccine should be designed specifically for each individual person according to his/her HLA alleles.
2. Each person's response to the virus infection might be correlated to his/her HLA alleles (in fact, this diversity is part of the natural selection of a species to survive). This hypothesis may have consequences in treatment of patients and prevention.
3. Generating such optimal CD8+ T cells may have therapeutic consequences, as early and rapid response of highly efficient T cells could help to control the virus multiplications hence preventing cytokine storm in the lung.

A similar analysis was performed on the nucleocapsid protein and the results are reported in Table 2. HLA class 1 has three loci A (68 common alleles), B (125 common alleles), C (44 common alleles), thus a total of 237 common alleles. We propose to extend Tables 1 and 2 from 20 alleles to all 237 common alleles. The predicted T-cell epitopes can be further verified by mass spectrometry and immune response assays.

Figure 2 gives two methods of applying the vaccine, modulo methods of selecting proper adjuvants such as dendritic cells. To design an epitope-based vaccine for each person, we first identify the HLA alleles. Then based on the allele-epitope associations, we can assemble an HLA-specific vaccine as depicted in Figure 2a. Given the current limited data (only the genome of the virus), our candidate peptides are predicted solely from the HLA alleles. When the immunopeptidomics data is available, we can train a personal deep learning model on the immunopeptidome of each individual person to improve the epitope prediction.

**Acknowledgements**

This work was funded in part by the NSERC OGP0046506 grant and the Canada Research Chair program.

**Competing Interests Statement**

The authors declare no competing interests.

**Figure Legends**

**Figure 1.** Workflow for obtaining data, training our prediction models, and obtaining neoantigens for each HLA allele.

**Figure 2.** Vaccination workflow. (a) Simple vaccination method. (b) Fully personalized vaccination.

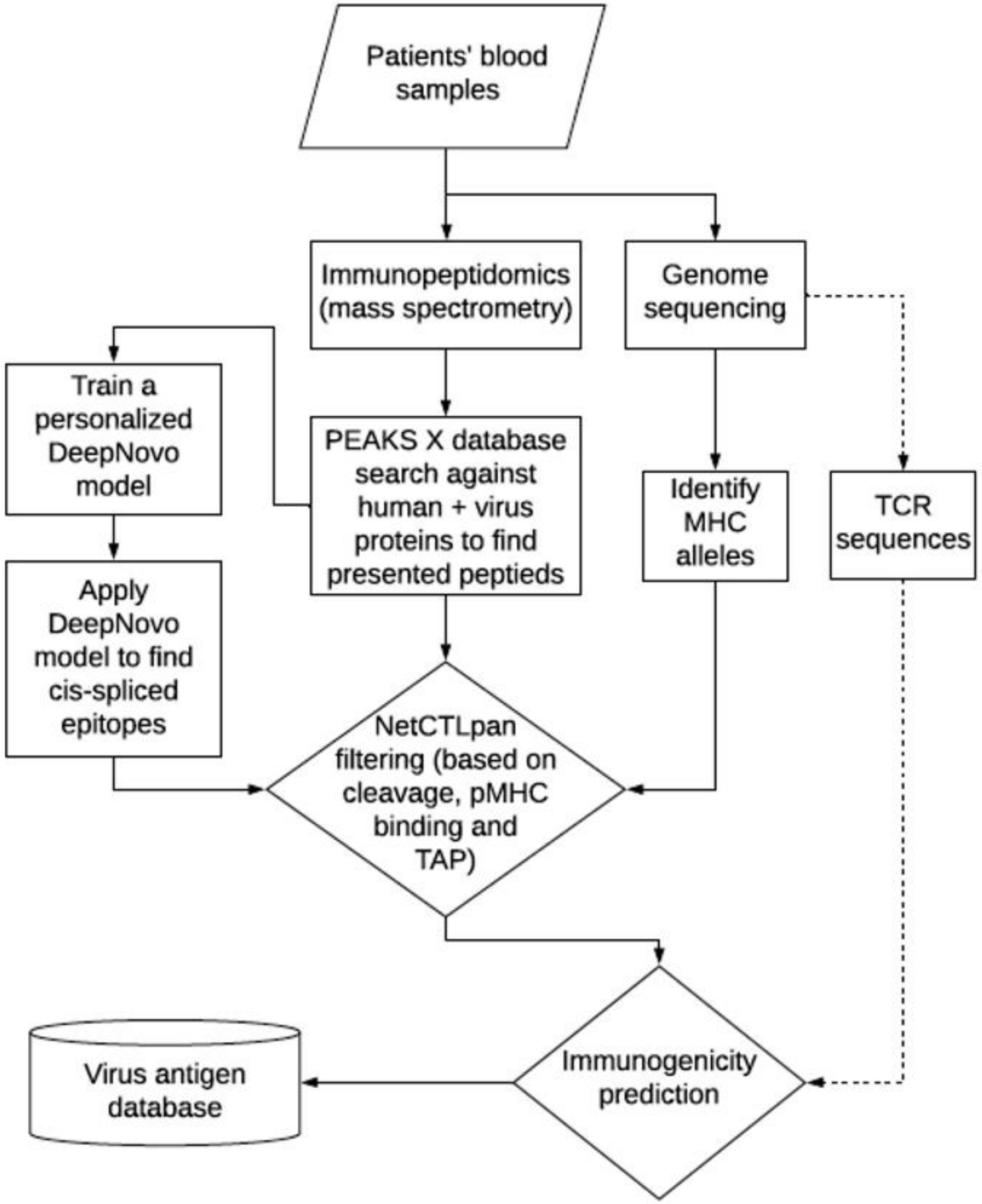
Patients' blood samples
Immunopeptidomics (mass spectrometry)
Genome sequencing
Train a personalized DeepNovo model
PEAKS X database search against human + virus proteins to find presented peptieds
Identify MHC alleles
TCR sequences
Apply DeepNovo model to find cis-spliced epitopes
NetCTLpan filtering (based on cleavage, pMHC binding and TAP)
Virus antigen database
Immunogenicity prediction

(a)

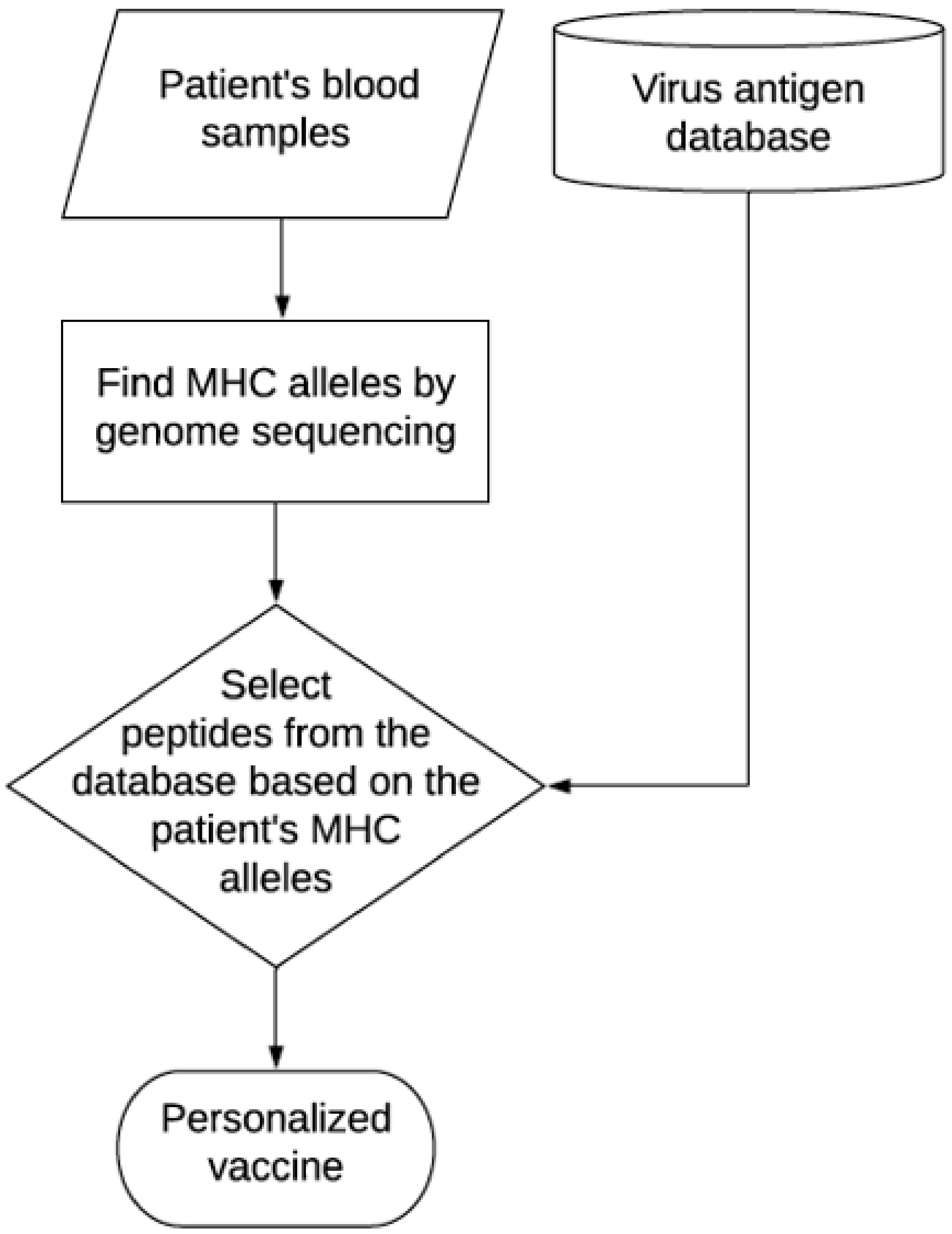
Patient's blood samples
Virus antigen database
Find MHC alleles by genome sequencing
Select peptides from the database based on the patient's MHC alleles
Personalized vaccine

(b)

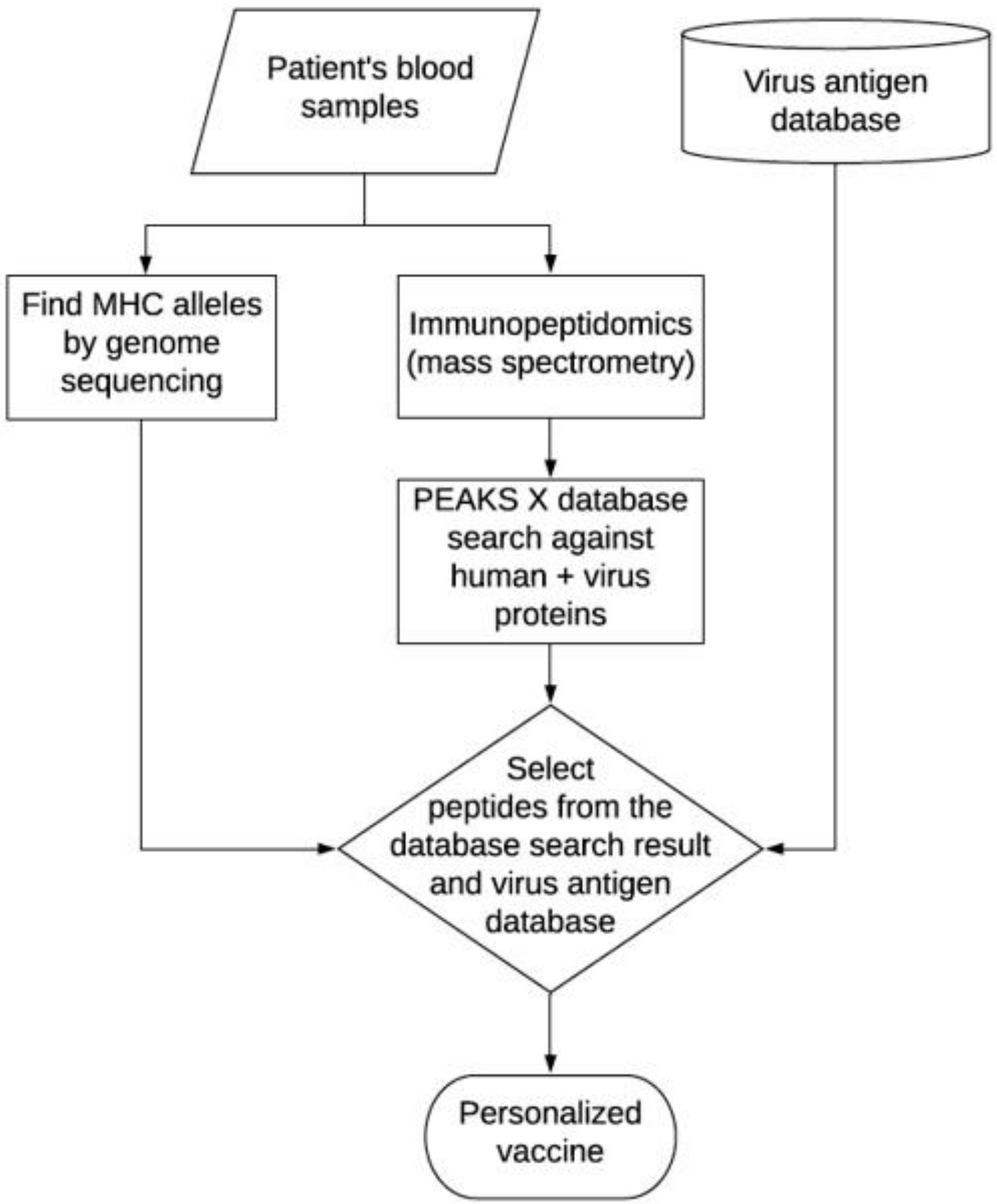
Patient's blood samples
Virus antigen database
Find MHC alleles by genome sequencing
Immunopeptidomics (mass spectrometry)
PEAKS X database search against human + virus proteins
Select peptides from the database search result and virus antigen database
Personalized vaccine

Table 1. Top 3 candidate peptides from S protein.

| HLA alleles | Allele frequency in Chinese population | Candidate 1 | Candidate 2 | Candidate 3 |
|---|---|---|---|---|
| HLA-A*11:01 | 0.277 | GVYFASTEK | ASANLAATK | VTYVPAQEK |
| HLA-A*24:02 | 0.172 | QYIKWPWYI | IAIPTNFTI | PYRVVVLSF |
| HLA-C*01:02 | 0.169 | FVFLVLLPL | RGWIFGTTL | IAIPTNFTI |
| HLA-B*40:01 | 0.149 | AEVQIDRLI | KEIDRLNEV | FERDISTEI |
| HLA-C*03:04 | 0.128 | WTFGAGAAL | FVFLVLLPL | FTISVTTEI |
| HLA-C*08:01 | 0.126 | WTFGAGAAL | FQFCNDPFL | IAIPTNFTI |
| HLA-A*33:03 | 0.115 | SVYAWNRKR | YNYLYRLFR | GTHWFVTQR |
| HLA-B*46:01 | 0.115 | FTISVTTEI | STQDLFLPF | FVSNGTHWF |
| HLA-A*02:03 | 0.108 | FIAGLIAIV | VLNDILSRL | VVFLHVTYV |
| HLA-A*02:07 | 0.094 | YLQPRTFLL | FIAGLIAIV | VVFLHVTYV |
| HLA-B*58:01 | 0.089 | RSFIEDLLF | LAGTITSGW | IAIPTNFTI |
| HLA-C*03:02 | 0.087 | STQDLFLPF | FVFLVLLPL | QSAPHGVVF |
| HLA-B*13:01 | 0.082 | YEQYIKWPW | IAIPTNFTI | FERDISTEI |
| HLA-B*15:02 | 0.071 | WMESEFRVY | LPFNDGVYF | QLTPTWRVY |
| HLA-B*38:02 | 0.071 | YQPYRVVVL | SQSIIAYTM | YLQPRTFLL |
| HLA-A*02:01 | 0.053 | YLQPRTFLL | FIAGLIAIV | FQFCNDPFL |
| HLA-B*51:01 | 0.05 | IPTNFTISV | LPFFSNVTW | IAIPTNFTI |
| HLA-B*15:01 | 0.044 | WMESEFRVY | STQDLFLPF | GTITSGWTF |
| HLA-C*04:01 | 0.044 | RGWIFGTTL | SNVTWFHAI | IAIPTNFTI |
| HLA-C*03:03 | 0.041 | WTFGAGAAL | FVFLVLLPL | FTISVTTEI |

Table 2. Top 3 candidate peptides from N protein.

| HLA alleles | Allele frequency in Chinese population | Candidate 1 | Candidate 2 | Candidate 3 |
|---|---|---|---|---|
| HLA-A*11:01 | 0.277 | KTFPPTEPK | LSPRWYFYY | |
| HLA-A*24:02 | 0.172 | YYRRATRRI | KHWPQIAQF | TWLTYTGAI |
| HLA-C*01:02 | 0.169 | SPRWYFYYL | NTASWFTAL | TPSGTWLTY |
| HLA-B*40:01 | 0.149 | | | |
| HLA-C*03:04 | 0.128 | NTASWFTAL | SPRWYFYYL | |
| HLA-C*08:01 | 0.126 | NTASWFTAL | SPRWYFYYL | LTYTGAIKL |
| HLA-A*33:03 | 0.115 | IGYYRRATR | NVTQAFGRR | GYYRRATRR |
| HLA-B*46:01 | 0.115 | NTASWFTAL | | |
| HLA-A*02:03 | 0.108 | GMSRIGMEV | | |
| HLA-A*02:07 | 0.094 | | | |
| HLA-B*58:01 | 0.089 | LSPRWYFYY | KHWPQIAQF | |
| HLA-C*03:02 | 0.087 | | | |
| HLA-B*13:01 | 0.082 | | | |
| HLA-B*15:02 | 0.071 | NTASWFTAL | TPSGTWLTY | |
| HLA-B*38:02 | 0.071 | KHWPQIAQF | | |
| HLA-A*02:01 | 0.053 | GMSRIGMEV | | |
| HLA-B*51:01 | 0.05 | NPANNAAIV | SPRWYFYYL | LPNNTASWF |
| HLA-B*15:01 | 0.044 | | | |
| HLA-C*04:01 | 0.044 | SPRWYFYYL | TWLTYTGAI | YYRRATRRI |
| HLA-C*03:03 | 0.041 | NTASWFTAL | SPRWYFYYL | |